\documentclass[manuscript]{aastex}
\usepackage{natbib}
\usepackage{epsfig}
\def\comma{,~}

\slugcomment{Version of 11 March 2005}
\shorttitle{Constraints on VHE gamma-ray emission from GRBs}
\shortauthors{Atkins et al.}

\begin{document}

\title{Constraints on Very High Energy gamma-ray emission from Gamma-Ray Bursts}

\author{
R.~Atkins,\altaffilmark{\ref{wisc},\ref{atkinscurrent}} 
W.~Benbow,\altaffilmark{\ref{ucsc},\ref{benbowcurrent}} 
D.~Berley,\altaffilmark{\ref{umcp}} 
E.~Blaufuss,\altaffilmark{\ref{umcp}} 
D.~G.~Coyne,\altaffilmark{\ref{ucsc}} 
T.~DeYoung,\altaffilmark{\ref{ucsc},\ref{umcp}} 
B.~L.~Dingus,\altaffilmark{\ref{lanl}} 
D.~E.~Dorfan,\altaffilmark{\ref{ucsc}} 
R.~W.~Ellsworth,\altaffilmark{\ref{georgemason}} 
L.~Fleysher,\altaffilmark{\ref{nyu}} 
R.~Fleysher,\altaffilmark{\ref{nyu}} 
M.~M.~Gonzalez,\altaffilmark{\ref{wisc}} 
J.~A.~Goodman,\altaffilmark{\ref{umcp}} 
E.~Hays,\altaffilmark{\ref{umcp},\ref{hayscurrent1},\ref{hayscurrent2}} 
C.~M.~Hoffman,\altaffilmark{\ref{lanl}} 
L.~A.~Kelley,\altaffilmark{\ref{ucsc}} 
C.~P.~Lansdell,\altaffilmark{\ref{umcp}}
J.~T.~Linnemann,\altaffilmark{\ref{msu}}
J.~E.~McEnery,\altaffilmark{\ref{wisc},\ref{mcenerycurrent}}
A.~I.~Mincer,\altaffilmark{\ref{nyu}} 
M.~F.~Morales,\altaffilmark{\ref{ucsc},\ref{moralescurrent}} 
P.~Nemethy,\altaffilmark{\ref{nyu}} 
D.~Noyes,\altaffilmark{\ref{umcp}} 
J.~M.~Ryan,\altaffilmark{\ref{unh}} 
F.~W.~Samuelson,\altaffilmark{\ref{fda}} 
P.~M.~Saz~Parkinson,\altaffilmark{\ref{ucsc}}
A.~Shoup,\altaffilmark{\ref{uci}} 
G.~Sinnis,\altaffilmark{\ref{lanl}} 
A.~J.~Smith,\altaffilmark{\ref{umcp}} 
G.~W.~Sullivan,\altaffilmark{\ref{umcp}} 
D.~A.~Williams,\altaffilmark{\ref{ucsc}}
M.~E.~Wilson,\altaffilmark{\ref{wisc}} 
X.~W.~Xu\altaffilmark{\ref{lanl}} 
and 
G.~B.~Yodh\altaffilmark{\ref{uci}}} 

\altaffiltext{1}{\label{wisc} Department of Physics, University of Wisconsin, 1150 University Ave, Madison, WI 53706}
\altaffiltext{2}{\label{atkinscurrent} Current address: Department of Physics, University of Utah, 115 South 1400 East, Salt Lake City, UT 84112}
\altaffiltext{3}{\label{ucsc} Santa Cruz Institute for Particle Physics, University of California, 1156 High Street, Santa Cruz, CA 95064}
\altaffiltext{4}{\label{benbowcurrent} Current address:  Max-Planck-Institut f\"ur Kernphysik, Postfach 103980, D-69029 Heidelberg, Germany}
\altaffiltext{5}{\label{umcp} Department of Physics, University of Maryland, College Park, MD 20742}
\altaffiltext{6}{\label{lanl} Group P-23, Los Alamos National Laboratory, P.O. Box 1663, Los Alamos, NM 87545}
\altaffiltext{7}{\label{georgemason} Department of Physics and Astronomy, George Mason University, 4400 University Drive, Fairfax, VA 22030}
\altaffiltext{8}{\label{nyu} Department of Physics, New York University, 4 Washington Place, New York, NY 10003}
\altaffiltext{9}{\label{hayscurrent1} Current address: High Energy Physics Division, Argonne National Laboratory, Argonne, IL 60439}
\altaffiltext{10}{\label{hayscurrent2} Current address: Enrico Fermi Institute, University of Chicago, Chicago, IL, 60637}
\altaffiltext{11}{\label{msu} Department of Physics and Astronomy, Michigan State University, 3245 BioMedical Physical Sciences Building, East Lansing, MI 48824}
\altaffiltext{12}{\label{mcenerycurrent} Current address: NASA Goddard Space Flight Center, Greenbelt, MD 20771}
\altaffiltext{13}{\label{moralescurrent} Current address: Massachusetts Institute of Technology, Building 37-664H, 77 Massachusetts Avenue, Cambridge, MA 02139}
\altaffiltext{14}{\label{unh} Department of Physics, University of New Hampshire, Morse Hall, Durham, NH 03824} 
\altaffiltext{15}{\label{fda} Office of Science and Engineering Laboratories, Center for Devices and Radiological Health, U.S. Food and Drug Administration.}
\altaffiltext{16}{\label{uci} Department of Physics and Astronomy, University of California, Irvine, CA 92697}

\begin{abstract}
The Milagro gamma-ray observatory employs a water Cherenkov detector to observe extensive 
air showers produced by high energy particles interacting in the Earth's atmosphere. 
Milagro has a wide field of view and high duty cycle, monitoring the northern sky almost 
continuously in the 100 GeV to 100 TeV energy range. Milagro is, thus, uniquely capable of searching 
for very high-energy emission from 
gamma-ray bursts (GRBs) during the prompt emission phase. Detection of $>$100 GeV counterparts 
would place powerful constraints on GRB mechanisms.
Twenty-five satellite-triggered GRBs occurred within the field of view of Milagro between
January 2000 and December 2001. We have searched for counterparts to
these GRBs and found no significant emission from any of the burst positions. 
Due to the absorption of high-energy gamma rays by the extragalactic background light, detections 
are only expected to be possible for redshifts less than $\sim$0.5.
Three of the GRBs studied have measured redshifts. GRB 010921 has a redshift
low enough (0.45) to allow an upper limit on the fluence to place an 
observational constraint on potential GRB models.
\end{abstract}


\section{Introduction} 

The search for prompt very high energy (VHE) emission ($>$100 GeV) from gamma-ray bursts (GRBs)
is motivated by both experimental observations and theoretical predictions, and its detection could
allow us to constrain GRB emission models. Although no observation has yet 
conclusively demonstrated VHE emission from any single burst, 
there have been several indications of emission at these high energies.

Milagrito, a prototype of Milagro, reported evidence for emission above 650 GeV from 
GRB 970417a, with a (post-trials) probability of 1.5$\times10^{-3}$ of being a background 
fluctuation~\citep{atkins00a,atkins03}. This search included 53 other bursts from which no significant 
emission was detected. The Tibet air shower array reported a correlation between $\sim$10 TeV air 
showers and a sample of 57 GRBs detected by the Burst And Transient Source Experiment 
(BATSE)~\citep{amenomori96}, although no significant signal was detected from any of the directions 
of the individual BATSE bursts. Evidence at about the 3 sigma level with HEGRA has been published 
for emission above 20 TeV from 
GRB 920925c~\citep{padilla98}. Follow-up observations above 250 GeV by the Whipple atmospheric Cherenkov 
telescope~\citep{connaughton97} did not find any high energy afterglow for 9 bursts studied.
Because this search involved slewing the Whipple telescope into position after
receiving a burst alert from BATSE (for these 9 bursts the delays ranged from 2 minutes 
to 56 minutes), only delayed emission could be detected. In addition, the small field 
of view of Whipple (3$^{\circ}$), compounded with the relatively large uncertainty in the BATSE 
location of the bursts ($\sim10^{\circ}$ in diameter) meant the process of searching the burst 
region required multiple pointings, taking a 3 hour observing cycle and still covering less 
than 50\% of the actual BATSE error box, further hampering any early detection of TeV 
emission~\citep{connaughton97}. 

Although most GRBs have been detected in the energy range
between 20 keV and 1 MeV, a few bursts have been observed at
energies above 100 MeV by EGRET, including the detection of 
an 18 GeV photon from GRB 940217, over 90 minutes after the start of the burst~\citep{hurley94}, indicating both that the spectra of some GRBs extend to at least GeV energies and that 
this emission may be delayed~\citep{dingus95,dingus01}.

Recently, a second spectral component was found in GRB 941017~\citep{gonzalez03}.
This second component extended up to at least 200 MeV, and its flux
decayed more slowly than the lower energy component.  
It is still unknown to how high in energy the second component extends. 
It is not clear whether it is similar to the high 
energy peak seen in TeV sources and attributed to inverse Compton emission
or the result of a completely different mechanism. The second
component has a very hard, power law spectrum, with a differential photon index of -1$\pm$0.3, which 
if extrapolated to 100 GeV would make the burst extremely bright, 
with a fluence greater than 10$^{-3}$ erg s$^{-1}$ cm$^{-2}$.

The observation of VHE emission from GRBs is hindered by collisions between the gamma rays 
and the extragalactic infrared background light (EBL), producing 
electron-positron pairs~\citep{nikishov61}. The degree of gamma-ray extinction from this 
effect is uncertain, because the amount of EBL is not well known. Direct measurements of 
the extragalactic background light have proven difficult, because of the foreground 
contribution from the Galaxy. There are several models of the extinction~\citep{primack99,stecker98,dejager02}, 
which are similar in their general features because of the constraints from the available data. Recent 
progress in the field has come about due to the accurate determination of the cosmological parameters, as 
well as a greatly increased knowledge of the luminosity function of galaxies. The most 
recent model now predicts a somewhat smaller absorption than was previously 
expected~\citep{primack04}, with an optical depth predicted to be roughly unity to 
500 GeV (10 TeV) gamma rays from a redshift of 0.2 (0.05).

While VHE emission from GRBs has been elusive, it is a natural byproduct of most GRB 
production models and is often predicted to have a fluence comparable to that at 
MeV energies~\citep{dermer00,pilla98,zhang01}.  
This is a result of the fact that the MeV emission from 
GRBs is likely synchrotron radiation produced by energetic electrons within the strong 
magnetic field of a jet with bulk Lorentz factors exceeding 100. In such an environment,
the inverse Compton mechanism for transferring energy from electrons to gamma rays is 
likely to produce a second higher energy component of GRB emission with fluence possibly
peaked at 1 TeV or above.  The relative strengths of the synchrotron and inverse Compton
emission depend on the environments of the particle acceleration and the gamma ray 
production.

In this paper, we use the Milagro gamma-ray observatory to search for VHE emission during the 
prompt emission of GRBs during 2000 and 2001. The GRBs have been detected by one or more 
of several satellite instruments: BATSE, Rossi X-ray Transient Explorer (RXTE), 
BeppoSax, High Energy Transient Explorer (HETE), and the Third Interplanetary Network 
(IPN), and occurred overhead within the field of view of Milagro. The combination 
of large field of view and high duty cycle make Milagro the best instrument available for conducting this type of search.

\section{The Milagro Observatory}

\begin{figure}[htbp]
\centering 
\rotatebox{-90}{\resizebox{!}{12cm}{
\includegraphics{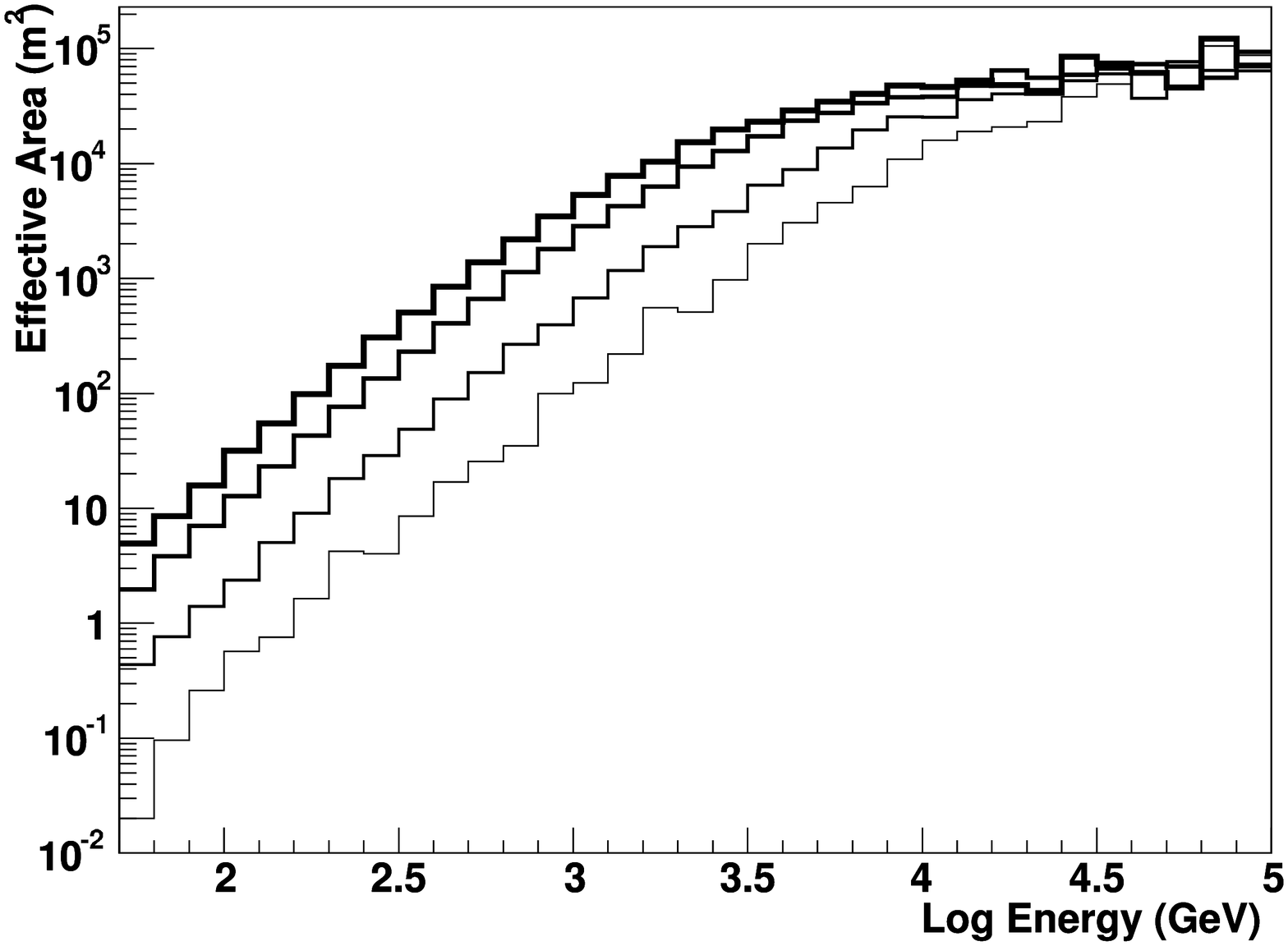}}}
\caption{Effective area of Milagro to gamma rays as a function of energy for various zenith angles. The different lines
reflect the effective area for different ranges of zenith angles (in decreasing order of thickness, 
0$^\circ$--15$^\circ$,15$^\circ$--30$^\circ$, 30$^\circ$--40$^\circ$, and 40$^\circ$--45$^\circ$) illustrating 
the drop suffered in effective area at energies below 1 TeV for large zenith angles.\label{fig1}}   
\end{figure}

Milagro is a TeV gamma-ray detector which uses the water Cherenkov technique to detect 
extensive air showers produced by very high-energy gamma rays as they traverse the Earth's 
atmosphere~\citep{atkins00b}. Milagro is located in the Jemez Mountains of northern 
New Mexico (35.9$^\circ$ N, 106.7$^\circ$ W) at an altitude of 2630 m above sea level,  
and has a field of view of $\sim$2 sr and a duty cycle of over 90\%, making it an 
ideal all-sky monitor of transient phenomena at very high energies, such as GRBs. 
The effective area and energy threshold of Milagro are a function of 
zenith angle, due to the increased atmospheric overburden at larger zenith 
angles, which tends to attenuate the particles in the air shower before they reach the ground. 
During the period covered by these observations, the Milagro trigger required that
approximately 55 tubes be hit within a 200 ns time window. For more details on Milagro 
see \cite{atkins03b}.
Figure~\ref{fig1} shows the effective area to gamma rays of Milagro during the years 
2000 and 2001. 
The 
plot shows the effective area for four different ranges of zenith angles:  
0$^\circ$--15$^\circ$,15$^\circ$--30$^\circ$, 30$^\circ$--40$^\circ$, and 40$^\circ$--45$^\circ$, illustrating 
the drop suffered in effective area at energies below a few TeV for large zenith angles. As can be seen, the 
sensitivity of Milagro varies slowly with angle out to around 30 degrees and then drops off. 
The effective area of Milagro ranges from $\sim$10 m$^{2}$ at 100 GeV to $\sim$10$^{5}$ m$^2$ at 10 TeV. Systematic effects result in an uncertainty in the effective area estimated to be no more than 15\%.
The angular resolution is approximately 0.7 degrees.

The energy response of Milagro is rather broad with no clear point to
define as an instrument threshold. To obtain a rough guide of the range of energies
to which Milagro is sensitive, we consider a power law spectrum 
with a differential photon index, $\alpha$, of -2.4. 
The energy (E$_{5}$) above which 95\% of the triggered events  
from such a spectrum are obtained is approximately 250 GeV, the energy (E$_{95}$) below which 95\% of the 
triggered events come is 25 TeV, and the median energy is 2.5 TeV. This illustrates both 
the breadth of the energy response of Milagro, showing that the Milagro Detector is sensitive 
to a low energy ($<$ 500 GeV) signal.

\section{The GRB sample}

Table~\ref{grb_table} shows a summary of the sample of satellite-triggered GRBs within the field of 
view (up to zenith angles of 45$^{\circ}$) of Milagro during 2000 and 2001. There were
three bursts during this two year period which were within the Milagro field of view but occurred 
when Milagro was not operating. The GRB sample presented here 
represents approximately a third of all well-localized bursts during that period \citep{greiner04}. 
Of the sample, three have known redshift, with GRB 010921 (z=0.45) being the closest. The detector 
configuration was upgraded starting in 2002 and later bursts will be the topic of a future paper.

The first column in Table~\ref{grb_table} is the GRB name, which, following the usual convention, represents
the UTC date (YYMMDD) on which the burst took place. The second column gives the instrument(s) that detected 
the burst. In the case of BATSE, we list the trigger ID. The third column gives the time at which
the burst triggered the particular instrument, in Truncated Julian Date (TJD)\footnote{TJD $\equiv$ Truncated 
Julian Date = JD - 2,440,000.5 = MJD - 40,000} followed by the UTC second
of the day. Column four gives the coordinates (right ascension and declination, in degrees) of the burst. The fifth 
column gives the uncertainty in the measured location of the burst in degrees. A dashed line implies that the 
error was small compared to the Milagro point-spread function (PSF). For BATSE bursts, the error radius is the 90\% 
systematic plus statistical uncertainty (added in quadrature). The duration listed in the sixth column is the  T90 interval for that burst as reported by the respective instrument teams. 
When times were reported for several energy bands, we picked the duration in the highest 
energy band. The seventh column gives the redshift of the GRB, when known. Column eight describes what (if any) afterglows were detected in different energy bands: (O)ptical, (R)adio, or (X)-ray. If an afterglow was observed, it is 
marked with a check mark, if it was not detected it is marked with a cross, and if no observation was attempted it is marked with a dot.
Column nine lists the zenith angle of the burst at Milagro, in degrees. We include only 
bursts for which the zenith angle was less than 45$^{\circ}$. As is implied by Figure~\ref{fig1}, the effective 
area of Milagro at zenith angles greater than 45$^{\circ}$ becomes negligible in the energy range where we 
expect GRB emission to be detectable (e.g. $<$ 1 TeV). The remaining columns of the table
list the Milagro results and are described later. First, we describe in a little more detail 
the most interesting GRB from the sample.

\subsection{GRB 010921}

GRB 010921 was detected by the HETE satellite~\citep{2002ApJ...571L.127R}, 
and together with data from the Interplanetary Network (IPN) it was localized 
to a 310-square-arcmin 3-sigma error box. It had a duration (T$_{90}$) of 24.6 seconds in 
the 7--400 keV energy range and a fluence in the 30--400 keV range of 
1.02$\times 10^{-5}$ erg cm$^{-2}$~\citep{Barraud03}. A variable optical source exhibiting 
a power law decay measured by the Large Format Camera on the Palomar 200-inch telescope was 
identified as the afterglow of GRB 010921~\citep{Price01}. This was superimposed on a bright 
galaxy, assumed to be the host galaxy, determined to be at z=0.45 \citep{Djorgovski01}. This 
was the first afterglow detected from a HETE-localized burst. 
Observations with the Hubble Space Telescope failed to detect a coincident SN to a 
limit 1.33 magnitude fainter than SN 1998bw at the 99.7\% confidence level, making this one of the most sensitive searches for an underlying SN \citep{Price03}.
A radio afterglow was detected by the VLA~\citep{Price01}. 
This is the first burst in the field of view of Milagro known to be close enough to be 
potentially detectable above 100 GeV.

The photon spectrum of GRB 010921 in the (7--200) keV range can be fit with a cut-off power 
law, defined by $\mathrm{dN/dE=AE^{\alpha}\exp(-E/E_0)}$ with $\alpha$ equal to $-1.49^{+0.7}_{-0.6}$, 
and $E_0$ equal to $206^{+81}_{-48}$ keV~\citep{Barraud03}. \cite{sakamoto} obtained slightly
different parameters ($\alpha = -1.55^{+0.08}_{-0.07}$, $E_0 = 197^{+48}_{-31}$ keV) 
by doing a joint spectral fit using WXM in addition to FREGATE data.
The data can also be fit by the Band function~\citep{band}, defined by 
$\mathrm{dN/dE=AE^{\alpha}\exp(-E/E_0)}$ for $E\leq(\alpha-\beta)E_0$, and $\mathrm{dN/dE=BE^{\beta}}$ 
for $E\geq(\alpha-\beta)E_0$, where 
$B=A[(\alpha-\beta)\times E_0]^{(\alpha-\beta)}\times\exp{(\beta-\alpha)}$. The best
fit parameters obtained from this fit were a low-energy power law index $\alpha$ equal to 
$-1.52^{+0.16}_{-0.09}$, a high-energy power law index $\beta$ equal to $-2.33^{+0.34}_{-7.67}$, 
and $E_0$ equal to $165^{+61}_{-59}$ keV (T. Q. Donaghy 2005, private communication). 
The break energy $E_0$ is related to the peak energy in the $\nu F_{\nu}$ spectrum by the 
equation $E_p=E_0\times(2+\alpha)$. The high-energy power law index, $\beta$, is not very well
constrained by the HETE data, as seen by the results of the fit. Figure~\ref{fig4b} shows the shape of the 
Band function with $\beta$ values equal to -1.99, -2.33, and -2.67. As $\beta$ becomes increasingly 
negative, the high-energy power law eventually becomes irrelevant in the HETE energy range as the 
fitting function reduces to the simple absorbed power law.
 
\section{Data Analysis}

\begin{figure}[htbp]
\centering 
\rotatebox{0}{\resizebox{!}{12cm}{
\includegraphics{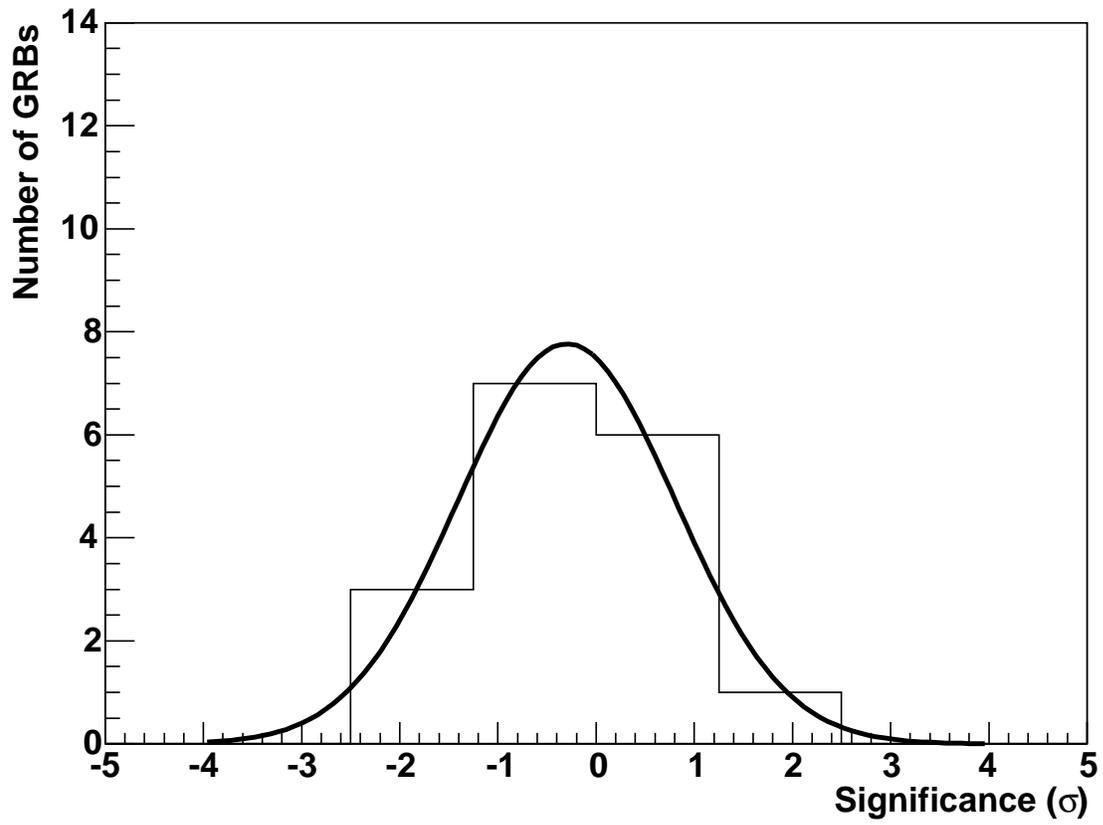}}}
\caption{Distribution of significances of the 17 well-localized GRBs.\label{fig2}}   
\end{figure}

A search for an excess of events above those expected from the background was
made for each of the 25 bursts in the sample. The search was performed for two different
durations, T90 (the time interval over which ninety per cent of the flux is observed) and 
five minutes. Although the value of T90 is derived from observations made at much lower 
energies than Milagro detects, EGRET showed that this duration is relevant at higher energies
too. Four GRBs which were observed with EGRET were among the five brightest 
bursts observed by BATSE and the significance of the EGRET detections in the T90 interval ranged 
from 6 to 12 sigma, leading to the speculation that all GRBs might have high energy emission 
during their resepctive T90 time intervals, and EGRET simply did not have the sensitivity to detect the rest of the
bursts~\citep{dingus01}. Figure~\ref{fig2} shows the distribution of significances of the
17 well-localized bursts for each T90 duration. The distribution plotted can be fit with a gaussian of 
mean -0.3 and standard deviation of 1.1. We also chose to search for emission from these GRBs for 
a duration of five
minutes. This timescale was motivated by the recent discovery of a second higher energy 
component in GRB 941017. While the T90 for that burst was 77 s, the second component 
(which has a fluence more than three times greater than the fluence in the BATSE energy range 
alone) has a duration of approximately 211 seconds~\citep{gonzalez03}.

The total number of events falling within a circular bin of radius $1.6^{\circ}$ at the 
location of the burst was summed for the duration of the burst and is shown in column 
ten ($\mathrm{N_{ON}}$) of Table~\ref{grb_table}. 
An estimate of the number of background events was 
made by characterizing the angular distribution of the background using two hours 
of data surrounding the burst~\citep{atkins03b}. This number is given as
$\mathrm{N_{OFF}}$ in column eleven of Table~\ref{grb_table}.
The significance of the excess (or deficit) for each burst was evaluated using 
equation [17] of~\cite{lima} and is given in column twelve. For bursts with an uncertainty in the 
position greater than 0.5$^{\circ}$, the search region (given by column five of 
Table~\ref{grb_table}) was tiled with an array of overlapping $1.6^{\circ}$ radius bins 
spaced $0.1^{\circ}$ apart in right ascension and declination and the numbers given in columns 
ten and eleven are chosen as the ones giving the excess with maximum significance. Since we do  
not take into account the total number of bins searched, this number does not represent the 
significance of detection of this burst and we therefore do not include these values with those
of the well-localized bursts, in column 12. No significant excess was found from any burst in the 
sample. The 99\% confidence upper limits on the number of signal events 
detected, $\mathrm{N_{UL}}$, given the observed $\mathrm{N_{ON}}$ and the predicted background
$\mathrm{N_{OFF}}$, following the method described by~\cite{helene} are given in column thirteen. 
Finally, we convert the upper limit on the counts into an upper limit on the fluence.
Using the effective area of Milagro, $\mathrm{A_{eff}}$, and assuming a differential power-law 
photon spectrum we integrate in the appropriate energy range and solve for the normalization 
constant. We chose a spectrum of the form $dN/dE=KE^{-2.4}$ photons/TeV/m$^2$. The power law index of 2.4 was chosen as a conservative value for the spectrum. The average spectrum of the four previously mentioned
bright bursts observed by EGRET had a differential photon spectrum with index 
1.95$\pm$0.25~\citep{dingus01} and the average spectrum of high energy blazars can be fit with a
simple power law with an index of 2.2~\citep{weekes}. 
The normalization factor K can be calculated by solving the following equation: 
$N_{UL}=\int{A_{eff}(dN/dE)dE}$.
Finally, we integrate the photon spectrum multiplied by the energy to obtain the corresponding 
value for the total fluence: $F=\int{E(dN/dE)dE}$, integrating from 0.25 to 25 TeV.

\section{Results}

Table~\ref{grb_table} shows a summary of GRBs between January 2000 and 
December 2001 analyzed with Milagro. None of these bursts showed any significant emission in
 the Milagro data.
The 99\% confidence upper limits on the fluence for both durations are listed in columns fourteen and 
fifteen of Table~\ref{grb_table} respectively. Note that none of the results listed in the table take 
into account the effect of absorption from the EBL. For GRB 000301C and GRB 000926, both
of which are at z$\sim$2, we expect essentially all of the TeV emission (if it exists) to be 
absorbed by the EBL, making the actual upper limit on the emission of TeV gamma rays from
these bursts considerably weaker than what is reflected in the table. For all the other bursts
(except GRB 010921) there are no redshift measurements, so it is not possible to take this effect
into account. The analysis of GRB 010921, which we have described in detail, yields an  
interesting upper limit, as we describe in the next section.

\subsection{GRB 010921}

\begin{figure}[htbp]
\centering 
\rotatebox{-0}{\resizebox{!}{12cm}{
\includegraphics{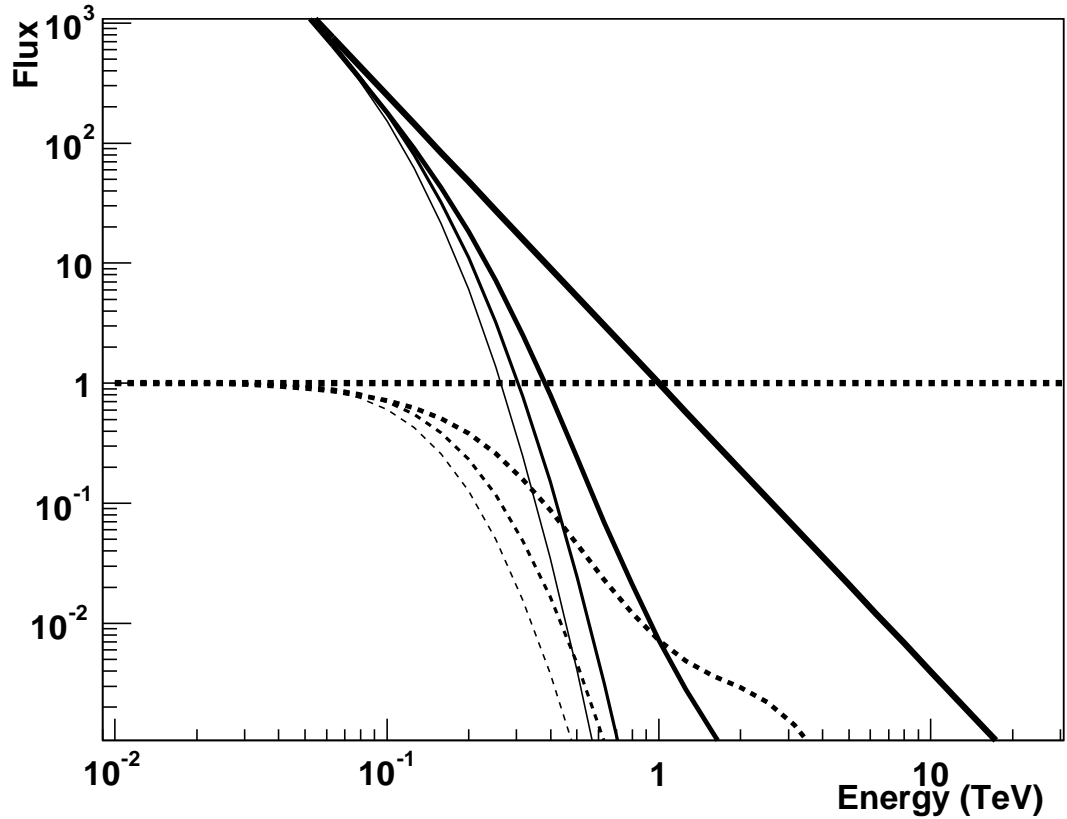}}}
\caption{Effect of the EBL on an E$^{-2.4}$ spectrum for z=0.45. The solid lines represent (in order
of decreasing thickness) the unabsorbed spectrum, the model of \cite{primack04}, and the Baseline and 
Fast Evolution models of \cite{stecker98}. The dashed lines represent the previous four curves divided
by the unabsorbed spectrum; i.e. the attenuation factor.\label{fig3}}   
\end{figure}

\begin{figure}[htbp]
\centering 
\rotatebox{0}{\resizebox{!}{10cm}{
\includegraphics{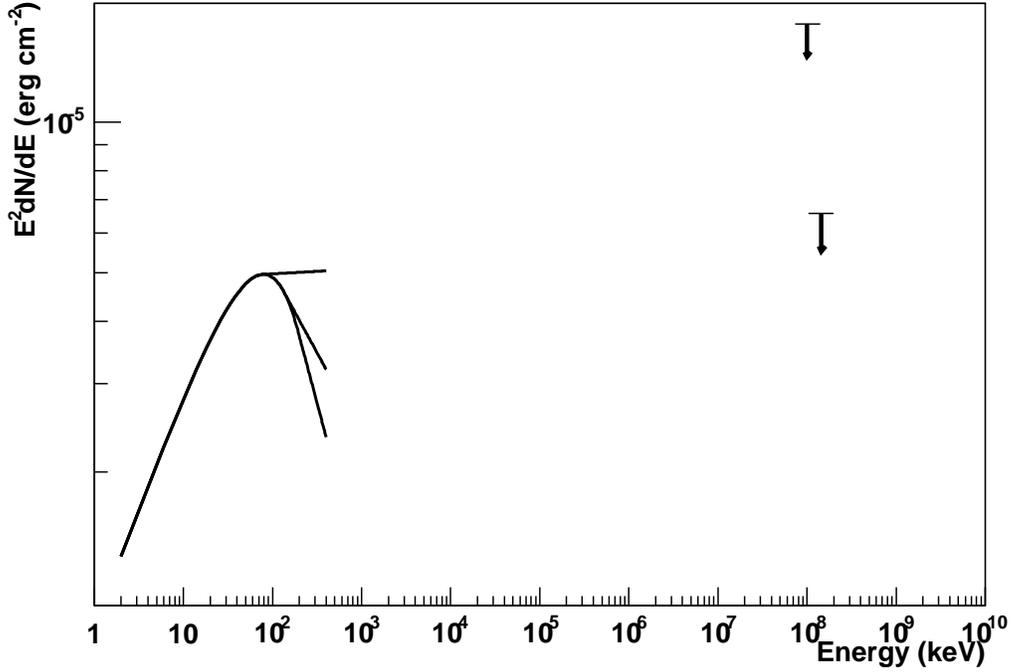}}}
\caption{The GRB 010921 spectrum as measured by HETE (shown in the range 2--400 keV). The function is the
result of fitting the data to the Band function. The high-energy power-law index $\beta$ is unconstrained. We
plot it for three values: -2.33, -1.99, and -2.67. The arrows are the upper 
limits from Milagro for various EBL absorption models, the lower energy one
corresponding to the \cite{stecker98} and \cite{primack99} models (the results are very similar) and the slightly
higher energy one corresponding to the more recent \cite{primack04} model. The energies at which these are 
evaluated are the respective median energies of the detected events and the quantity plotted 
is $E^2dN/dE$.\label{fig4b}}   
\end{figure}

GRB 010921, with a measured redshift of z=0.45, is close enough that a potential TeV signal
is not totally absorbed by the extragalactic background. 
Figure~\ref{fig3} shows the effect of various absorption models at a redshift of 0.45 on an 
E$^{-2.4}$ spectrum which might be expected from a GRB. 
For an unbroken power law, the energy range from 0.25 TeV to 25 TeV is the sensitive region for Milagro.
For a cutoff spectrum, the small amount of sensitivity between 50 and 250 GeV is not negligible, so we
now include it.
Integrating the product of a normalization constant K times an E$^{-2.4}$ energy spectrum multiplied 
by the EBL absorption and effective area in the energy range 0.05--100 TeV and 
setting it equal to the 99\% confidence upper limit on the counts (20.2), allows us to solve for K 
(2.0$\times10^{-2}$ photons TeV$^{-1}$ m$^{-2}$). 
We then use this normalization constant to compute the unabsorbed fluence.
The upper limits obtained are $<2.9\times10^{-5}$ erg cm$^{-2}$ for the \cite{primack04} model,  
$5.8\times10^{-5}$ erg cm$^{-2}$ for the \cite{primack99} model, and $<5.8\times10^{-5}$ erg cm$^{-2}$
for the ``fast evolution'' model of \cite{stecker98} (the more absorptive of their two models).
Figure~\ref{fig4b} shows the spectrum of GRB 010921 obtained by HETE (in the energy range of 7 to 200 
keV, where the data were fitted) as well as the Milagro upper limits. The quantity plotted is 
E$^2$dN/dE, and has been evaluated at the median energy of events that would be detected resulting 
from using each of the EBL absorption models, approximately 100 GeV for both the ``fast evolution'' 
model of \cite{stecker98} and the \cite{primack99} model, and 150 GeV for the \cite{primack04} model. 
The derived upper limits of E$^2$dN/dE, using these three absorption models are 1.6$\times10^{-5}$erg 
cm$^{-2}$ (for the first two models) and 6.7$\times10^{-6}$erg cm$^{-2}$ (for third one). These are 
shown as two separate arrows in Figure~\ref{fig4b}.

\section{Discussion}

The conclusive detection of TeV emission from GRBs would improve our current understanding not
only of GRBs themselves, but also of a whole range of physical phenomena. Besides possibly allowing
us to learn something about the magnetic fields and electron energies in the GRB environments, as in 
the case of blazars, it could also serve to distinguish particle acceleration models, constrain
the models of the infrared photon density of the universe, and possibly to probe some more exotic
phenomena such as quantum gravity~\citep{amelino}.
The number of GRBs visible to Milagro so far has been relatively small. The fraction of GRBs expected 
to be detected by Milagro depends on many assumptions. As mentioned in the introduction, there are good theoretical
as well as experimental reasons to expect GRBs to emit photons at GeV to TeV energies. 
Even if these photons were created at (and escaped) the source, however, one of the greatest 
limitations in their detection involves the attenuation due to the intergalactic background radiation.
High energy $\gamma$-rays interact with intergalactic star light, creating electron 
positron pairs~\citep{primack99,stecker98,dejager02,primack04}. The attenuation 
is a function of gamma-ray energy, the density and spectrum of the 
background radiation fields and, crucially, the distance to the source.
Even at the relatively modest redshift of 0.5, the optical depth is greater than 1 at a few hundred GeV.
Figure~\ref{fig3} shows the effect of the EBL on an E$^{-2.4}$ spectrum 
(shown, unabsorbed, as a straight line) at a redshift of 0.45, according to different models.
Unfortunately, only three of the 25 GRBs studied here have a measured redshift, and of these, only
GRB 010921, with a redshift of 0.45, was close enough for a significant fraction of photons to escape
absorption. 

\begin{figure}[htbp]
\centering 
\rotatebox{0}{\resizebox{!}{10cm}{
\includegraphics{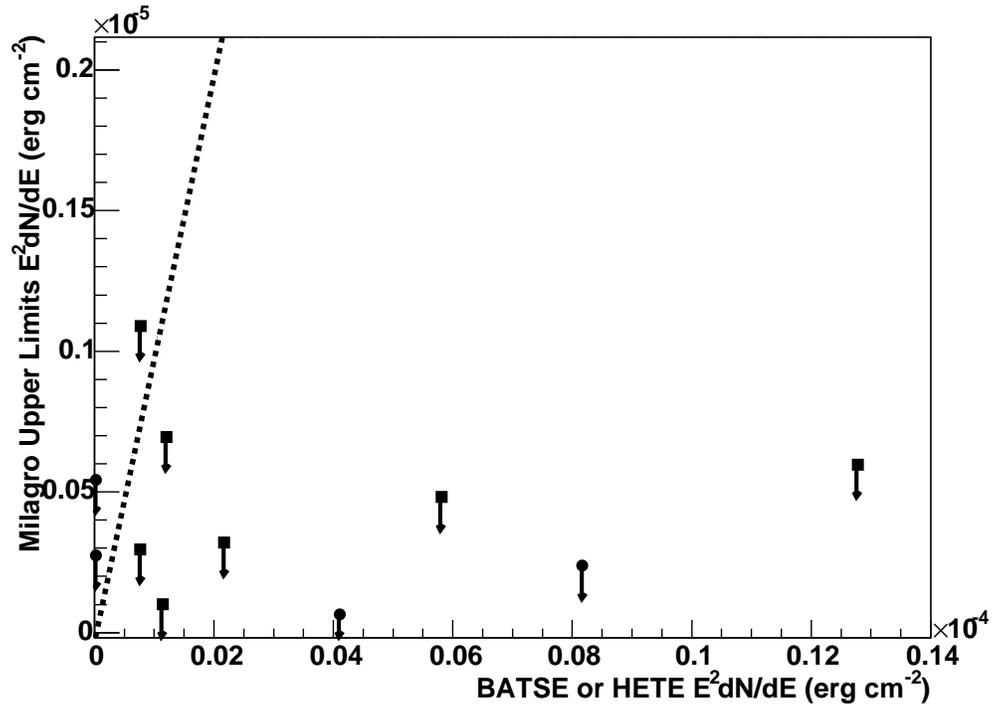}}}
\caption{Comparison of BATSE (squares) and HETE (circles) fluences to Milagro upper limits. The bursts
represented in this plot are: GRB 000113, 000212, 000226, 000302, 000317, 000331, 000508, 010613, 
010921, 011130, and 011212. The dotted line represents the line y=x.\label{fig5}}   
\end{figure}

Figure~\ref{fig5} compares the Milagro upper limits for a number of bursts with the observations made 
by BATSE and HETE (we show those for which a fluence in the lower energy bands was available). The 
quantity plotted is E$^{2}$dN/dE. We have assumed 
an E$^{-2}$ spectrum at the lower energies and an E$^{-2.4}$ spectrum for the Milagro energy range, 
evaluating it at 2.5 TeV. The circles represent HETE bursts while squares represent BATSE bursts. 
It is important to note that the absorption from the EBL has not been taken into account. Because 
the redshifts of these bursts are unknown, the Milagro upper limits cannot place tight constraints
on the burst emission. For many bursts, the Milagro fluence limits are below those of HETE or BATSE,
so that if the burst were known to be close, the limit would be constraining.
GRB 010921 was the first HETE burst with a measured afterglow. The VHE limit from Milagro is the first which 
constrains the TeV flux rather than the distance to the source.

\begin{figure}[htbp]
\centering 
\rotatebox{-90}{\resizebox{!}{12cm}{
\includegraphics{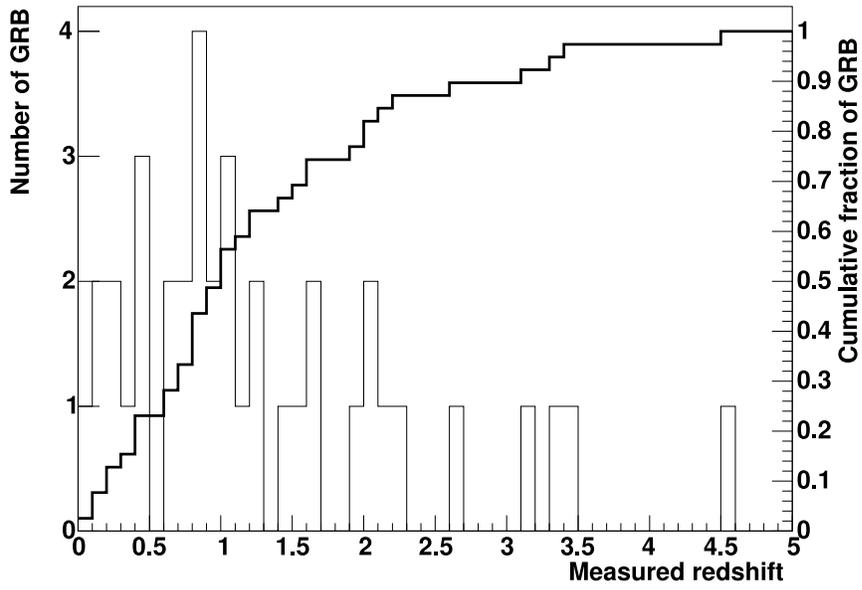}}}
\caption{Distribution of the 39 currently measured redshifts, including bursts GRB 970228 
through GRB 041006. The scale on the right refers to the cumulative distribution also
shown in this plot.
\label{fig6}}   
\end{figure}

With the launch of Swift, we expect $\sim$50 well-localized bursts
in the Milagro field of view during the first two years, most of them with measured redshifts. 
There are currently 39 GRBs with known redshift~\citep{greiner04}. Figure~\ref{fig6} shows a histogram
of these redshifts (left scale), with the cumulative distribution also shown on the plot (right scale). 
Approximately 20\% of the measured redshifts are 0.5 or less.
A sample of known low redshift GRBs, combined with the upgrades made to Milagro, increasing the effective area 
of Milagro and lowering its energy threshold, should allow us in the near future to improve greatly on 
the current results and shed some light on the nature of VHE emission from GRBs.

\section{Conclusions}

A search for very high energy emission from GRBs was performed with the
Milagro observatory in the range of 50 GeV to 100 TeV. A total of 25 
satellite-triggered GRBs were within the field of view of Milagro in the 
two year period between January 2000 and December 2001, including GRB 010921,
at a known redshift of 0.45. No significant emission was detected from any
of these bursts. 99\% confidence upper limits on the fluence are presented.

\begin{acknowledgments}
Many people helped bring Milagro to fruition.  In particular, we
acknowledge the efforts of Scott DeLay, Neil Thompson and Michael Schneider. We have 
used GCN Notices to select raw data for archiving and use in this search, and
we are grateful for the hard work of the GCN team, especially Scott Barthelmy.
We thank Tim Donaghy and the rest of the HETE team for helpful comments as well
as access to HETE data for GRB010921.
This work has been supported by the National Science Foundation (under grants 
PHY-0075326, 
-0096256, 
-0097315, 
-0206656, 
-0245143, 
-0245234, 
-0302000, 
and
ATM-0002744) 
the US Department of Energy (Office of High-Energy Physics and 
Office of Nuclear Physics), Los Alamos National Laboratory, the University of
California, and the Institute of Geophysics and Planetary Physics.
\end{acknowledgments}

\newpage

\newpage
\pagestyle{empty}
\begin{deluxetable}{lllllllllllllll}
\rotate
\renewcommand{\footnoterule}{\rule{0pt}{0pt}}

\tabletypesize{\tiny}
\tablewidth{0pt}
\tablecaption{List of GRBs in the field of view of Milagro in 2000 and 2001\label{grb_table}}

\tablehead{
\colhead{GRB} & 
\colhead{ID\tablenotemark{2}} &
\colhead{Time\tablenotemark{3}} &
\colhead{Pos.\tablenotemark{4}} & 
\colhead{Err. circ.\tablenotemark{5}} & 
\colhead{T90/Dur.(s)} & 
\colhead{z\tablenotemark{7}}	& 
\colhead{ORX\tablenotemark{8}} &
\colhead{$\theta$\tablenotemark{9}} & 
\colhead{$\mathrm{N_{ON}}$\tablenotemark{10}} &
\colhead{$\mathrm{N_{OFF}}$\tablenotemark{11}} & 
\colhead{$\sigma$\tablenotemark{12}} &
\colhead{$\mathrm{N_{UL}}$\tablenotemark{13}} & 
\colhead{UL (T90/Dur.)\tablenotemark{14}} &
\colhead{UL (5 min)\tablenotemark{15}}
}

\startdata
000113 & 7948   & 11556:34202.4 & 163.27,+19.89	& 5.0  	& 370  	& \nodata & \nodata	& 20.9	& 471 	& 423.2 & \nodata & 95.8 & 5.5e-6 & 4.8e-6 \\
000212 & 7987   & 11586:81065.2 & 16.09,+35.62	& 4.2  	& 8 	& \nodata & \nodata	& 2.21	& 32    & 17.7	& \nodata & 30.1 & 1.1e-6 & 5.0e-6 \\
000226 & 7998   & 11600:13909.4 & 143.68,+29.82	& 4.5  	& 10   	& \nodata  & \nodata& 31.5	& 25    & 11.4  & \nodata & 27.9 & 3.4e-6 & 1.1e-5 \\
000301C& 8005R/I& 11604:35497 	& 245.09,+29.42 &  -   	& 14   	& 2.03 	& $\surd\surd$.	& 37.6	& 11	& 10.3 & 0.2 & 12.0 & 2.6e-6 & 6.6e-6 \\
000302 & 8008   & 11605:10225.1 & 58.20,+54.28	& 4.15 	& 120	& \nodata & \nodata	& 31.9	& 143 	& 113.1	& \nodata & 54.7  & 6.8e-6 & 1.2e-5\\
000317 & 8039   & 11620:77953.8 & 27.22,+32.66	& 5.5  	& 550	& \nodata & \nodata	& 6.39	& 1740	&1625.9 & \nodata & 208.0 & 7.9e-6 & 5.4e-6 \\
000330 & 8057   & 11633:75449.4 & 358.31,+39.26 & 5.4 	&$0.2^*$& \nodata	&\nodata	& 30.0	& 3 	& 0.3	& \nodata & 9.8 & 1.0e-6 & 1.2e-5\\
000331 & 8061   & 11634:85421.8 & 32.00,+59.77	& 5.9	& 55	& \nodata	& \nodata	& 38.3	& 94 	& 66.2	& \nodata & 53.0 & 1.2e-5 & 2.1e-5\\
000408 & I/8069 & 11642:9348.2 	& 137.32,+66.58 & 0.5	& 2.5	& \nodata &$\times$..& 31.1& 4 	& 3.4 	& 0.3 & 8.6 & 1.0e-6 & 6.7e-6\\
000508 & 8099   & 11672:77419.3 & 89.87,+2.41	& 4.3	& 30	& \nodata	& \nodata	& 34.1& 25 	& 15.3	& \nodata & 24.0  & 3.7e-6 & 9.4e-6\\
000615 & S      & 11710:22704	& 233.14,73.79 	& -	& 10	& \nodata	&$\times\times\times$& 39.0& 3 & 7.2 	& -1.8 & 6.3  & 1.6e-6 & 8.0e-6\\
000630 & I      & 11725:1853	& 221.81,+41.22 & -	& 20	& \nodata	&$\surd\times$.& 33.2	& 23 	& 22.6	& 0.1 & 15.5  & 2.2e-6 & 6.7e-6 \\
000727 & I      & 11752:70956	& 176.00,+17.41 & -	& 10	& \nodata	&$\times\times$.& 40.8	& 5  	& 6.2 	& -0.5 &  8.2  & 2.6e-6 & 1.2e-5\\
000730 & I      & 11755:255	& 191.29,+19.27 & -	& 7 	& \nodata	& \nodata	& 19.2& 8  	& 13.9	& -1.7 & 7.9 & 4.2e-7 & 2.7e-6 \\
000926 & I      & 11813:85773   & 256.06,+51.78 & -	& 25	& 2.04	&$\surd\surd\surd$& 15.9 & 60 	& 56.8	& 0.4 & 25.2 & 1.2e-6 & 3.4e-6\\
001017 & I      & 11834:80346	& 272.18,-2.99	& 0.5	& 10	& \nodata	& \nodata	& 42.1	& 2 	& 4.6	& -1.3 & 6.1 & 2.2e-6 & 1.0e-5\\
001018 & I      & 11835:61114	& 198.54,+11.81 & 0.5	& 31	& \nodata	& .$\surd$.& 31.8	& 31 	& 31.5	& -0.1 & 16.9 & 2.1e-6 & 6.8e-6\\
001019 & I      & 11836:86375	& 257.93,+35.34 & -	& 10	& \nodata	& $\times$..& 19.5	&  27	& 21.1 	& 1.2 & 20.9  & 1.1e-6 & 1.1e-6\\
001105 & I      & 11853:59128   & 195.3,+35.49	& -	& 30	& \nodata	& $\times$..& 8.5	&  87	& 76.2	& 1.2 & 35.6 & 1.4e-6 & 2.0e-6\\
010104 & I      & 11913:62489 	& 267.4,+18.23  & 0.5	& 2	& \nodata	& \nodata	& 19.8	&  3	& 3.3	& -0.2 & 7.5 & 4.0e-7 & 2.8e-6\\
010220 & S      & 11960:82267	& 39.4,+61.7	& -	& 150	& \nodata & $\times\times$.	& 27.0	& 155 & 168.8& -1.1 & 24.9 & 2.1e-6 & 3.0e-6\\
010613 & I/H	& 12073:27234	& 255.18,+14.27 & - 	& 152	& \nodata	& \nodata	& 24.7	& 277  	& 280.5 & -0.2 & 40.7 & 2.9e-6 & 5.1e-6\\
010921 & I/H	& 12173:18950.6 & 344.0,+40.93 	& -	& 24.6	&  0.45	&$\surd\surd$.& 10.4	& 61 	& 65.4	& -0.6 & 20.2 & 8.0e-7 & 2.2e-6\\
011130 & H      & 12243:22775.7 & 46.4,+3.8	& -	& 83.2	& \nodata &$\times\times\times$& 33.7 & 93 & 100.9 & -0.8 & 23.0 & 3.4e-6 & 8.3e-6\\
011212 & H      &  12255:14642 	& 75.051,+32.13 & -	& 84.4	& \nodata & \nodata& 33.0	& 132 	& 113.1 & 1.7 & 43.8 & 6.7e-6 & 1.2e-5\\


\tablecomments{
2) BATSE ID, when known, otherwise: R -- \emph{RXTE}, I -- IPN, H -- \emph{HETE}, S -- \emph{BeppoSAX};
3) Time of burst, TJD: seconds;
4) ra,dec in degrees;
5) Error circle in degrees (a dash implies error is small compared to the Milagro PSF);
7) redshift (when known);
8) afterglow observed: O -- Optical/IR, R -- Radio, X -- X-ray;
9) Zenith angle, in degrees;
10) Milagro number of counts on source;
11) Estimated number of background counts;
12) Statistical significance of the excess (in standard deviations);
13) 99\% Upper Limit on the number of counts coming from the source;
14) 99\% Upper Limit on the fluence (0.25--25 TeV), in ergs cm$^{-2}$ for the duration in column 6;
15) 99\% Upper Limit on the fluence (0.25--25 TeV), in ergs cm$^{-2}$ for a duration T = 5 min. 
*) Estimate. 
}
\enddata

\end{deluxetable}

\normalsize


\begin{thebibliography}{10}
\bibliographystyle{plain}

\bibitem[Amenomori et al.(1996)]{amenomori96} Amenomori, M. et al. 1996, \aap, 311, 919

\bibitem[Amelino-Camelia et al.(1998)]{amelino} Amelino-Camelia, G. et al. 1998, Nature, 393, 763

\bibitem[Atkins et al.(2000a)]{atkins00a} Atkins, R. et al. 2000a, \apjl, 533, L119

\bibitem[Atkins et al.(2000b)]{atkins00b} Atkins, R. et al. 2000b, Nucl.\ Instrum.\ Meth.\ A, 449, 478

\bibitem[Atkins et al.(2003a)]{atkins03} Atkins, R. et al. 2003a, \apj, 583, 824

\bibitem[Atkins et al.(2003b)]{atkins03b} Atkins, R. et al. 2003b, \apj, 595, 803

\bibitem[Atkins et al.(2004)]{atkins04} Atkins, R. et al. 2004, \apj, 608, 680

\bibitem[Atteia(2003)]{2003astro.ph..12371A} Atteia, J. 2003, {\tt astro-ph/0312371} 

\bibitem[Band et al.(1993)]{band} Band, D., et al.\ 1993, 
\apj, 413, 281 

\bibitem[Barraud et al.(2003)]{Barraud03} Barraud, C. et al. 2003, \aap, 400, 1021

\bibitem[Castro et al.(2000)]{Castro00} Castro, S.~M., et al. 2000, GCN 605

\bibitem[Castro et al.(2003)]{2003ApJ...586..128C} Castro, S., Galama, 
T.~J., Harrison, F.~A., Holtzman, J.~A., Bloom, J.~S., Djorgovski, S.~G., 
\& Kulkarni, S.~R.\ 2003, \apj, 586, 128 

\bibitem[Connaughton et al.(1997)]{connaughton97} Connaughton, V., et al. 1997, \apj, 479, 859 

\bibitem[Costa et al.(1997)]{1997Natur.387..783C} Costa, E., et al.\ 1997, 
\nat, 387, 783 

\bibitem[Dall et al.(2000)]{Dall00} Dall, T., et al. 2000, GCN Circ. 804

\bibitem[De Jager \& Stecker(2002)]{dejager02} de Jager, O.~C. \& Stecker, F.~W. 2002, \apj, 566,738

\bibitem[De Pasquale et al.(2003)]{2003ApJ...592.1018D} De Pasquale, M., et 
al.\ 2003, \apj, 592, 1018 

\bibitem [Dermer\comma Chiang \& Mitman(2000)]{dermer00} 
Dermer, C.\ D., Chiang, J., \& Mitman, K.\ E.\ 2000, \apj, 537, 785

\bibitem[Dingus(1995)]{dingus95} Dingus, B. L., 1995, Astrophys. \& Space Sci., 231, 187

\bibitem[Dingus(2001)]{dingus01}
Dingus, B. L., 2001, in \emph{High Energy Gamma Ray Astronomy}, ed. F. A. Aharonian and H. J. Volk, 2001, AIP, 558, 383 

\bibitem[Dingus(2003)]{dingus03}
Dingus, B. L., 2003, in \emph{Gamma-Ray Bursts: 30 Years of Discovery}, ed. E.~E.~Fenimore and M.~Galassi, 2003, AIP, 727, 131

\bibitem[Djorgovski et al.(2001)]{Djorgovski01} Djorgovski, S.~G., et al. 2001, GCN 1108

\bibitem[Fynbo et al.(2001)]{2001A&A...369..373F} Fynbo, J.~U.~et al.\ 2001, \aap, 369, 373 

\bibitem[Galama et al.(1998)]{1998Natur.395..670G} Galama, T.~J., et al.\ 
1998, \nat, 395, 670 

\bibitem[Gonzalez et al.(2003)]{gonzalez03} Gonzalez, M.~M.\ et al. 2003, Nature, 424, 749

\bibitem[Gorosabel et al.(2000)]{Gorosabel00} Gorosabel, J., et al. 2000, GCN Circ. 803

\bibitem[Greiner(2004)]{greiner04} Greiner, J. 2004, {\tt http://www.mpe.mpg.de/\~\,jcg/grbgen.html}

\bibitem[Helene(1983)]{helene} Helene, O. 1983, Nucl.\ Instrum.\ Methods Phys. Res., 212, 319

\bibitem[Hurley et al.(2000)]{Hurley00} Hurley, K., et al. 2000, GCN Circ. 736

\bibitem[Hurley et al.(1994)]{hurley94} Hurley, K., et al. 1994, \nat, 372, 652

\bibitem[Kulkarni et al.(1998)]{1998Natur.395..663K} Kulkarni, S.~R., et 
al.\ 1998, \nat, 395, 663 

\bibitem[Li \& Ma(1983)]{lima} Li, T.\ P., \& Ma, Y.\ Q.\ 1983, \apj, 272, 317

\bibitem[Maiorano et al.(2003)]{2003astro.ph..2022M} Maiorano, E.~et al.\ 2003, {\tt astro-ph/0302022} 

\bibitem[Nikishov (1961)]{nikishov61} Nikishov, A. I. 1961, Zh.\ Eksp.\ i Teor.\ Fiz., 41, 549 (English translation:Sov. Phys. JETP 14, 392 [1962])

\bibitem[Padilla et al.(1998)]{padilla98} Padilla, L., et al. 1998, \aap, 337, 43 

\bibitem[Pilla \& Loeb(1998)]{pilla98} Pilla, R. P. \& Loeb, A. 1998, \apjl, 494, L167 

\bibitem[Piro(2000)]{Piro00} Piro, L. 2000, GCN Circ. 703

\bibitem[Preece et al.(1998)]{Preece} Preece, R.~D., et al. 1998, \apjl, 506, L23

\bibitem[Price et al.(2001)]{Price01} Price, P.~A., et al. 2001, GCN Circ. 1107

\bibitem[Price et al.(2003)]{Price03} Price, P.~A., et al. 2003, \apj, 584, 931

\bibitem[Primack et al.(1999)]{primack99} Primack, J. R., Bullock, J. S., Somerville, R. S. \& Macminn, D.  1999, 
Astroparticle Physics, 11, 93 

\bibitem[Primack et al.(2004)]{primack04} Primack, J. R., Bullock, J. S., \& Somerville, R. S. 2004 
in \emph{Gamma 2004 Heidelberg}, ed. Aharonian, F.~A. 2004

\bibitem[Ricker et al.(2002)]{2002ApJ...571L.127R} Ricker, G.~, et al. 2002, \apjl, 571, L127 

\bibitem[Smith et al.(2001)]{Smith01} Smith, I.~A., Tilanus, R.~P.~J., Wijers, R.~A.~M.~J., 
Tanvir, N., Vreeswijk, P., Rol, E. \& Kouveliotou, C. 2001, \aap, 380,81

\bibitem[Sakamoto et al.(2004)]{sakamoto} Sakamoto, T., et al.\ 
2004, ArXiv Astrophysics e-prints, astro-ph/0409128

\bibitem[Stecker \& de Jager(1998)]{stecker98} Stecker, F. \& de Jager, O. C. 1998, \aap, 334, L85
 
\bibitem[van Paradijs et al.(2000)]{2000ARA&A..38..379V} van Paradijs, J.,  Kouveliotou, C. 
\& Wijers, R.~A.~M.~J. 2000, \araa, 38, 379

\bibitem[Weekes, T. C.(2003)]{weekes} Weekes, T.~C.~ 2003,\emph{Very High Energy Gamma-Ray Astronomy} (Bristol and Philadelphia: Institute of Physics)

\bibitem[Zhang \& M\'esz\'aros(2001)]{zhang01} Zhang, B.\ \& M\'esz\'aros, P.\ 2001, \apj, 559, 110

\end{thebibliography}
\end{document}